\documentstyle[preprint,eqsecnum,aps,floats,epsfig]{revtex}
\def\jtvec#1{\mbox{\boldmath $#1$}}

\def\yresult{\left(-0.5\pm 1.0{}^{+0.7}_{-0.8}\right)\times10^{-2}}
\def\yllimit{-0.030}
\def\yulimit{0.020}
\def\intlycp{23.4}
\newcommand{\jtprl}[3]{\prl {\bf #1}, #2 (#3)}
\newcommand{\jtplb}[3]{\pl B {\bf #1}, #2 (#3)}
\newcommand{\jtnpb}[3]{Nucl.\ Phys.\ {\bf B#1}, #2 (#3)}
\def\Yresult{$\ycp=\yresult$}

\def\Ylimit{$\yllimit<\ycp<\yulimit$}
\def\etal{{\it et al.}}
\def\ycp{y_{CP}}
\def\Fb{fb$^{-1}$}
\def\DzDzbar{$D^0$-$\overline{D}{}^0$}

\def\dz{D^0}
\def\dzbar{\overline{D}{}^0}
\def\Dz{$\dz$}

\def\pip{\pi^+}
\def\km{K^-}
\def\kp{K^+}
\def\kmkp{\km\kp}
\def\kpi{\km\pip}
\def\dzkpi{\dz\to\kpi}
\def\dzkk{\dz\to\kmkp}
\def\Gevc{GeV/$c$}
\def\Mevcsq{MeV/$c^2$}
\def\pt{p_T}

\def\tsig{\tau_{\rm SIG}}
\def\tbg{\tau_{\rm BG}}
\def\smis{S_{\rm tail}}
\def\smisbg{S_{\rm tail}^{\rm BG}}
\def\sbg{S_{\rm BG}}
\def\fmis{f_{\rm tail}}
\def\fmisbg{f_{\rm tail}^{\rm BG}}
\def\ftbg{f_{\tau_{\rm BG}}}
\def\eg{{\it e.g.}}
\begin{document}
\draft
\epsfysize3cm
\epsfbox{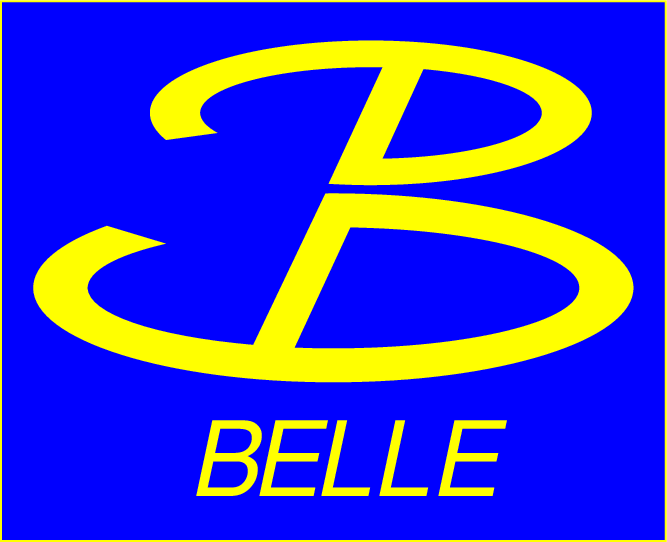}    
\vskip -3cm
\noindent
\hspace*{4.5in}KEK preprint 2001-143
\vskip -3mm
\noindent
\hspace*{4.5in}BELLE preprint 2001-18

\begin{center}
\vskip 2cm
{\Large \bf A Measurement of Lifetime Difference \\ in $D^0$ Meson Decays}
\\
The Belle Collaboration
\\
(November 8, 2001)
\end{center}
\tighten






\begin{abstract}
We report a measurement of the \DzDzbar\ mixing parameter $\ycp$
using \intlycp~\Fb\ of data collected
near the $\Upsilon$(4$S$) resonance
with the Belle detector at KEKB.
$\ycp$ is measured from the lifetime difference of \Dz\ mesons
decaying into the $K^-\pi^+$ state and the $CP$ even eigenstate $K^-K^+$.
We find \Yresult, where the first error is statistical and the second systematic,
corresponding to a 95\% confidence interval \Ylimit.
\end{abstract}
\pacs{PACS numbers: 12.15.Ff,13.25.Ft,14.40.Lb}

\begin{center}
  K.~Abe$^{8}$,               
  K.~Abe$^{42}$,              
  R.~Abe$^{31}$,              
  T.~Abe$^{43}$,              
  I.~Adachi$^{8}$,            
  Byoung~Sup~Ahn$^{16}$,      
  H.~Aihara$^{44}$,           
  M.~Akatsu$^{24}$,           
  Y.~Asano$^{49}$,            
  T.~Aso$^{48}$
% ...
\end{center}


\small
\begin{center}
% ...
$^{8}${High Energy Accelerator Research Organization (KEK), Tsukuba}\\
% ...
$^{12}${Institute of High Energy Physics, Vienna}\\
$^{13}${Institute for Theoretical and Experimental Physics, Moscow}\\
$^{14}${J. Stefan Institute, Ljubljana}\\
$^{15}${Kanagawa University, Yokohama}\\
$^{16}${Korea University, Seoul}\\
% ...
$^{24}${Nagoya University, Nagoya}\\
$^{25}${Nara Women's University, Nara}\\
% ...
$^{31}${Niigata University, Niigata}\\
% ...
$^{40}${University of Sydney, Sydney NSW}\\
% ...
$^{42}${Tohoku Gakuin University, Tagajo}\\
$^{43}${Tohoku University, Sendai}\\
$^{44}${University of Tokyo, Tokyo}\\
% ...
$^{48}${Toyama National College of Maritime Technology, Toyama}\\
$^{49}${University of Tsukuba, Tsukuba}\\
% ...
\end{center}





\newpage

The search for \DzDzbar\ mixing provides an important window on physics 
beyond the Standard Model~(SM).
Since the predicted rate of mixing in the SM is very small~\cite{mix-expect},
large mixing could indicate a contribution from non-SM processes.
One measure of mixing effects is the lifetime difference  between $D^0$ decaying to
the $K^- K^+$ final state (a $CP$-even eigenstate) and 
the $K^- \pi^+$ final state (which is not a $CP$ eigenstate)
$$\ycp \equiv \frac{\tau(K^- \pi^+)} { \tau(K^- K^+)} -1,$$ 
where $\tau = 1 / \hat{\Gamma}$
and $\hat{\Gamma}$ is the effective decay rate obtained by fitting a single
exponential to the measured decay distribution for each final
state~\cite{bias-exp}.
We combine decays from $\dz$ and $\dzbar$, which
we assume to be produced at equal rates in $e^+ e^-$ collisions.
The parameter $\ycp$ can be approximated as
$\ycp  \sim  y\cos\phi + x \Delta \sin\phi$.
Here $x = (M_1 - M_2) / \Gamma_{\rm av}$ and
$y = (\Gamma_1 - \Gamma_2) / 2\Gamma_{\rm av}$, where
$M_{1,2}$ and $\Gamma_{1,2}$ are the masses and decay widths for the
$|D_{1,2}\rangle =  p |\dz\rangle \pm q |\dzbar\rangle$ physical states of
the neutral $D$ meson system, and
$\Gamma_{\rm av} = \frac{1}{2} (\Gamma_1 + \Gamma_2)$.
$\phi$ is the phase of
$q  {\mathcal A}(\dzbar \rightarrow K^- K^+) /
          p  {\mathcal A}(\dz    \rightarrow K^- K^+)$, 
where ${\mathcal A}(\dz~(\dzbar) \rightarrow K^- K^+)$ are the decay amplitudes,
and
$\Delta = (|p|^2 - |q|^2)/(|p|^2 + |q|^2)$.
In the $CP$-conserving limit, $\ycp = y$~\cite{mix-th}.


The FOCUS collaboration recently reported
$y_{CP} = (3.42\pm 1.39\pm 0.74)\times10^{-2}$~\cite{FOCUS}.
A common SM prediction is $x, y \sim O(10^{-3})$.
Since non-SM processes are expected to enhance $x$ but not $y$,
such a large value of $\ycp$ would be difficult to interpret as a signal of new physics.
Possible SM effects at the $10^{-2}$ level would be a more natural explanation~\cite{bigi}.
We note however that limits on the mixing parameter $x$ are weak
($|x| < 0.03 \sim 0.06$), and the parameter $\Delta$ is not constrained
by existing measurements~\cite{CLEO}, 
so that a large $\ycp$ could also be accommodated
if $CP$ violating effects ($\Delta$ and/or $\phi$) were large.

In this Letter, we present a new measurement of $\ycp$ 
with better statistical precision than the FOCUS result and largely independent systematic errors.
The data sample for this analysis corresponds to an integrated luminosity
of 20.9~\Fb\ taken at the $\Upsilon$(4$S$) resonance and
of 2.5~\Fb\ taken 60~MeV below the $\Upsilon$(4$S$) resonance
collected with the Belle detector at the KEKB collider.
KEKB~\cite{kekb} is an asymmetric energy
electron-positron collider designed to produce boosted $B$ mesons.
The electron and positron beam energies are 8~GeV and 3.5~GeV,
respectively: the resulting energy in the center-of-mass system (cms), 10.58~GeV,
corresponds to the mass of the $\Upsilon$(4$S$) resonance.

The Belle detector~\cite{NIM} consists of
a three-layer double-sided~($r\phi$ and $rz$ planes) silicon vertex detector~(SVD),
a 50-layer central drift chamber~(CDC),
an array of 1188 aerogel \v{C}erenkov counters~(ACC),
128 time-of-flight~(TOF) scintillation counters,
an electromagnetic calorimeter containing 8736 CsI(Tl) crystals
and 14 layers of 4.7-cm-thick iron plates interleaved with a system of resistive plate counters~(KLM).
All subdetectors except the KLM are located inside
a 3.4 m diameter superconducting solenoid which provides a
1.5 Tesla magnetic field.
The impact parameter resolutions for charged tracks are measured  to be
$\sigma_{xy}^2=(19)^2 + (50/(p\beta\sin^{3/2}\theta))^2~\mu$m$^2$ in the plane perpendicular to the beam ($z$) axis and
$\sigma_{z}^2=(36)^2 + (42/(p\beta\sin^{5/2}\theta))^2~\mu$m$^2$ along the $z$ axis, 
where $\beta=pc/E$, $p$ and $E$ are  the momentum (GeV/$c$) and energy (GeV), and
$\theta$ is the polar angle from the $z$ axis.
The transverse momentum resolution is
$(\sigma_{\pt}/\pt)^2=(0.0019\pt)^2+(0.0030)^2$, where $\pt$ is in \Gevc.

\begin{figure}
\begin{center}
\resizebox{0.40\textwidth}{!}{\includegraphics{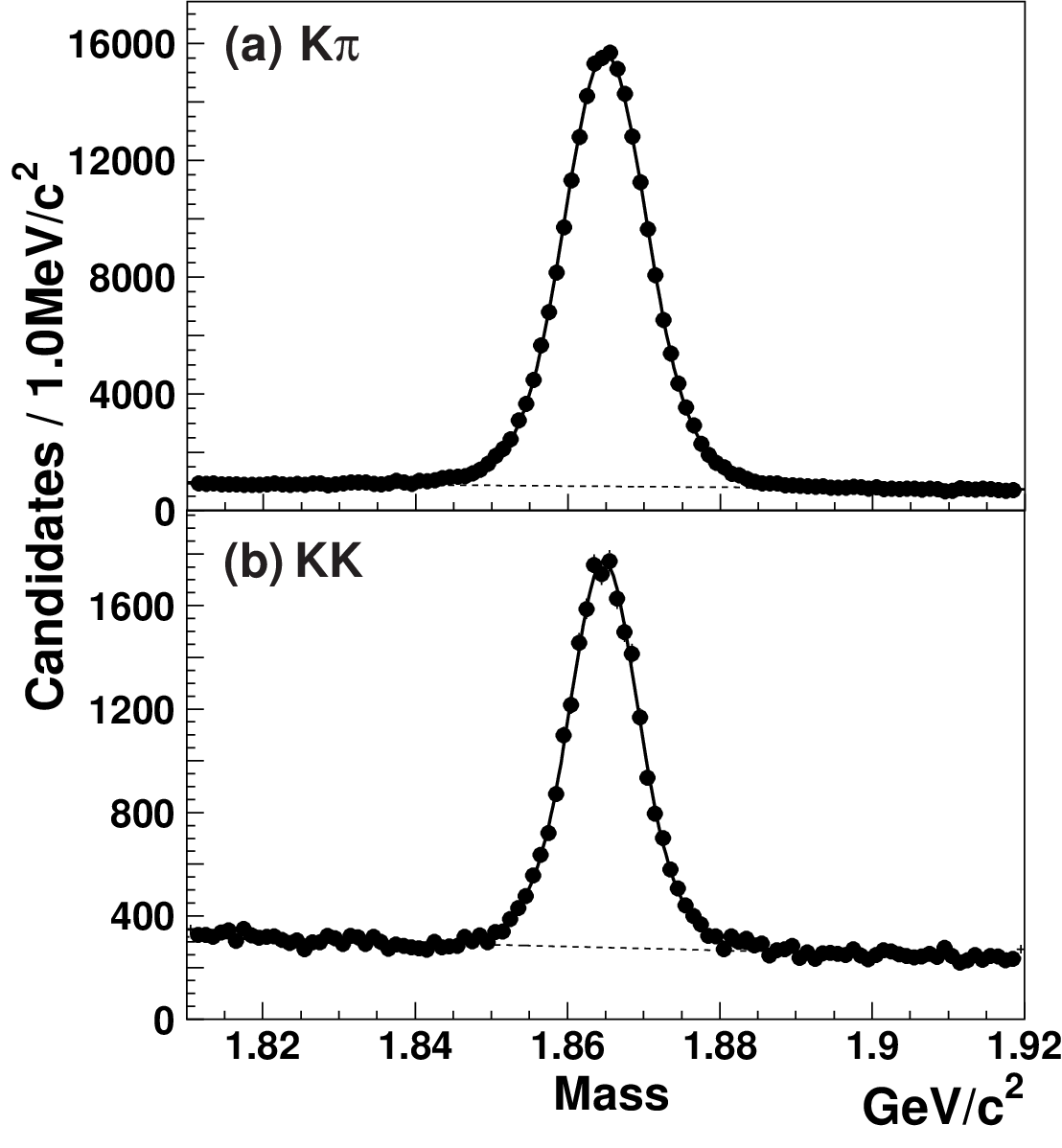}}
\end{center}
\begin{center}
\caption{The invariant mass distributions for (a) $\dzkpi$ and (b) $\dzkk$ candidates.
The results of the fit with two Gaussians~(signal) 
and a linear function~(background) are superimposed. 
The dotted line indicates the background.}
\label{fig:mass_distributions}
\end{center}
\end{figure}

Candidate \Dz\ mesons~\cite{CC} are reconstructed via 
$\dz\to\kpi$ and
$\dz\to\kmkp$ decays.
We require candidate charged tracks
to be well reconstructed, and
associated with at least two SVD hits in both the $r\phi$ and $rz$ planes.
Charged pion and kaon candidates are selected based on a particle identification likelihood value
calculated using CDC energy loss measurements, flight times measured in the TOF, and
the response of the ACC;
for the chosen cuts, $K^{\pm}$ are identified with              
an efficiency of $\sim$85\% and a charged pion fake                        
rate of $\sim$10\% for momenta up to 3.5~GeV/$c$.
A $D^0$ candidate is required to have cms momentum greater than 2.5 \Gevc\
to eliminate secondary $D^0$ mesons originating from $B\bar{B}$ events.
To suppress the background due to random combinations of two oppositely charged tracks, 
a cut is imposed on the decay angle $\theta_{\mathrm{D}}$ defined as
the angle between the momentum vector of the $D^0$ candidate in the laboratory frame
and that  of the $\pi^+$ ($K^+$) in the $D^0$ rest frame for the $K^-\pi^+$ ($K^-K^+$)  decay:
$\cos\theta_{\mathrm{D}}>-0.85$ for the $K^-\pi^+$ decay and
$|\cos\theta_{\mathrm{D}}|<0.90$ for the $K^-K^+$ decay,
where the $\cos\theta_{\mathrm{D}}$ distribution is expected to be flat for signal. 
The two tracks forming the $D^0$ candidate are required to originate from a common vertex.
In addition,  the reconstructed $D^0$ flight path is required to be consistent with originating at the $e^+e^-$ interaction point~(IP) profile.
Figure~\ref{fig:mass_distributions} shows invariant mass distributions for $\dzkpi$ and $\dzkk$ candidates,
superimposed with the result of a fit with two Gaussians (for signal) plus a linear function (for background).
We find $214260 \pm 562$ $\dzkpi$ and $18306 \pm 189$ $\dzkk$ signal events,
as determined from the area of the two Gaussians.
The signal purity in the mass region within $3\sigma$ of the measured mean $D^0$ mass 
is 87\% for $K^-\pi^+$  (67\% for $K^-K^+$), where $\sigma$ is the weighted average of the standard deviations of two Gaussians
and is $6.5~{\rm MeV}/c^2$ ($5.4~{\rm MeV}/c^2$ for $K^-K^+$). 

The decay vertex~($\jtvec{x}_{\rm dec}$) of the charm meson is determined using
both tracks that form the charm meson candidate.
The production vertex~($\jtvec{x}_{\rm pro}$) is obtained by extrapolating the $D^0$
flight path to the IP.
The center and size of the IP profile are
determined for each KEKB fill\cite{IP}.
The size of the IP region is
$\sigma_x=(80-120)~\mu$m,
$\sigma_y=(2-4)~\mu$m,  and
$\sigma_z=(3-4)$~mm.
The signed decay length  in 3-dimensional space~\cite{sign} and the proper-time~($t$) are obtained from
$\ell = (\jtvec{x}_{\rm dec}-\jtvec{x}_{\rm pro})\cdot\jtvec{p}_D/|\jtvec{p}_D|$ and
$t = \ell m_D/|\jtvec{p}_D|$, respectively, where $\jtvec{p}_D$ and $m_D$
are the momentum vector of the reconstructed charm meson and the world average value of the $D^0$ mass~\cite{pdg}.
The selected $D^0$ mesons have a laboratory momentum of $3.6~{\rm GeV}/c$ and 
a decay length of $\sim$200~$\mu$m on average.
A Monte Carlo (MC) simulation study indicates that the decay and production vertex resolutions 
along the $D^0$ flight direction are 110~$\mu$m (rms) and 70$~\mu$m (rms), respectively.
We reject a small fraction~($\sim$ 3\%) of the $D^0$ candidates
by requiring that the uncertainty of the decay length measurement be less than $300\ \mu$m.

In this analysis we extract the value of $\ycp$ by combining likelihood
functions for $\dz \rightarrow K^- \pi^+$ and $\dz \rightarrow K^- K^+$ decays,
$$
  L_{\ycp} = L_{K^- \pi^+} \cdot L_{K^- K^+},
$$
and expressing the lifetime for $\dz \rightarrow K^- K^+$ as
\[
  \tau(K^- K^+)  =  \tau(K^- \pi^+) / (1 + \ycp) 
\]
in an unbinned maximum likelihood fit.
This method allows us to properly estimate correlated systematic 
errors.
The likelihood functions $L = L_{K^- \pi^+}$ and $L_{K^- K^+}$ are given by~\cite{Lifetime}
\begin{eqnarray*}
&&L(\tsig, S, \smis, \fmis, \tbg, \ftbg, \sbg, \smisbg, \fmisbg)\\
&&\;= \prod_{i} \bigg[f_{\rm SIG}^i\int^\infty_0 \!\!\!\!\! dt^\prime\frac{1}{\tau_{\rm SIG}}e^{\frac{-t^\prime}
{\tau_{\rm SIG}}}
R(t^i-t^\prime;\sigma_t^i,S,\smis,\fmis) \\
&&\quad\quad\quad+
(1-f_{\rm SIG}^i)\int^\infty_0  \!\!\!\!\! dt^\prime
\{f_{\tau_{\rm BG}}\frac{1}{\tau_{\rm BG}}e^{\frac{-t^\prime}{\tau_{\rm BG}}} \\
&&\quad\quad\quad+(1-f_{\tau_{\rm BG}})\delta(t^\prime)\}
R(t^i-t^\prime;\sigma_t^i,\sbg,\smisbg,\fmisbg)\bigg],
\end{eqnarray*}
with separate parameters for $K^-\pi^+$ and $K^-K^+$,
giving a total of 18  parameters to fit  including $\ycp$.
The product is over the $D^0$ candidates.
Here, $f_{\rm SIG}^i$ is the probability that the candidate is signal,
calculated as a function of the invariant mass, and $\tau_{\rm SIG}$ is the signal lifetime.
The function $R$ represents the resolution of the proper time $t^i$.
The background proper-time distribution is modeled by
a fraction $\ftbg$~($\sim$ 0.15 for $K^-\pi^+$ and $\sim$ 0.21 for $K^-K^+$)
with the effective lifetime $\tbg$~($\sim$ 391~fs for $K^-\pi^+$ and $\sim$ 497~fs for $K^-K^+$)
and a fraction with zero lifetime
represented by the Dirac delta function $\delta(t^\prime)$.

The resolution function $R$, separately for the signal and the background, is parameterized as:
\begin{eqnarray*}
R(t;\sigma_t,S,S_{\rm tail},f_{\rm tail}) &=& (1-f_{\rm tail})\frac{1}{\sqrt{2\pi}S\sigma_t}
e^{-\frac{t^2}{2S^2\sigma_t^2}} \\
&+&f_{\rm tail}\frac{1}{\sqrt{2\pi}S_{\rm tail}\sigma_t}
e^{-\frac{t^2}{2S_{\rm tail}^2\sigma_t^2}},
\end{eqnarray*}
where $\sigma_t$ is the proper-time error estimated candidate-by-candidate from the decay length error, 
and $S$ and $S_{\rm tail}$ are global scaling factors that modify $\sigma_t$
for the main and tail Gaussian distributions and $f_{\rm tail}$ is the fraction of the tail part.
The main component is due to the SVD vertex resolution, while the tail component is
due to poorly reconstructed tracks~(\eg\ tracks affected by
misassociation of SVD hits,
wrong SVD hit clustering or large angle multiple-scattering).

\begin{figure}
\begin{center}
\resizebox{0.40\textwidth}{!}{\includegraphics{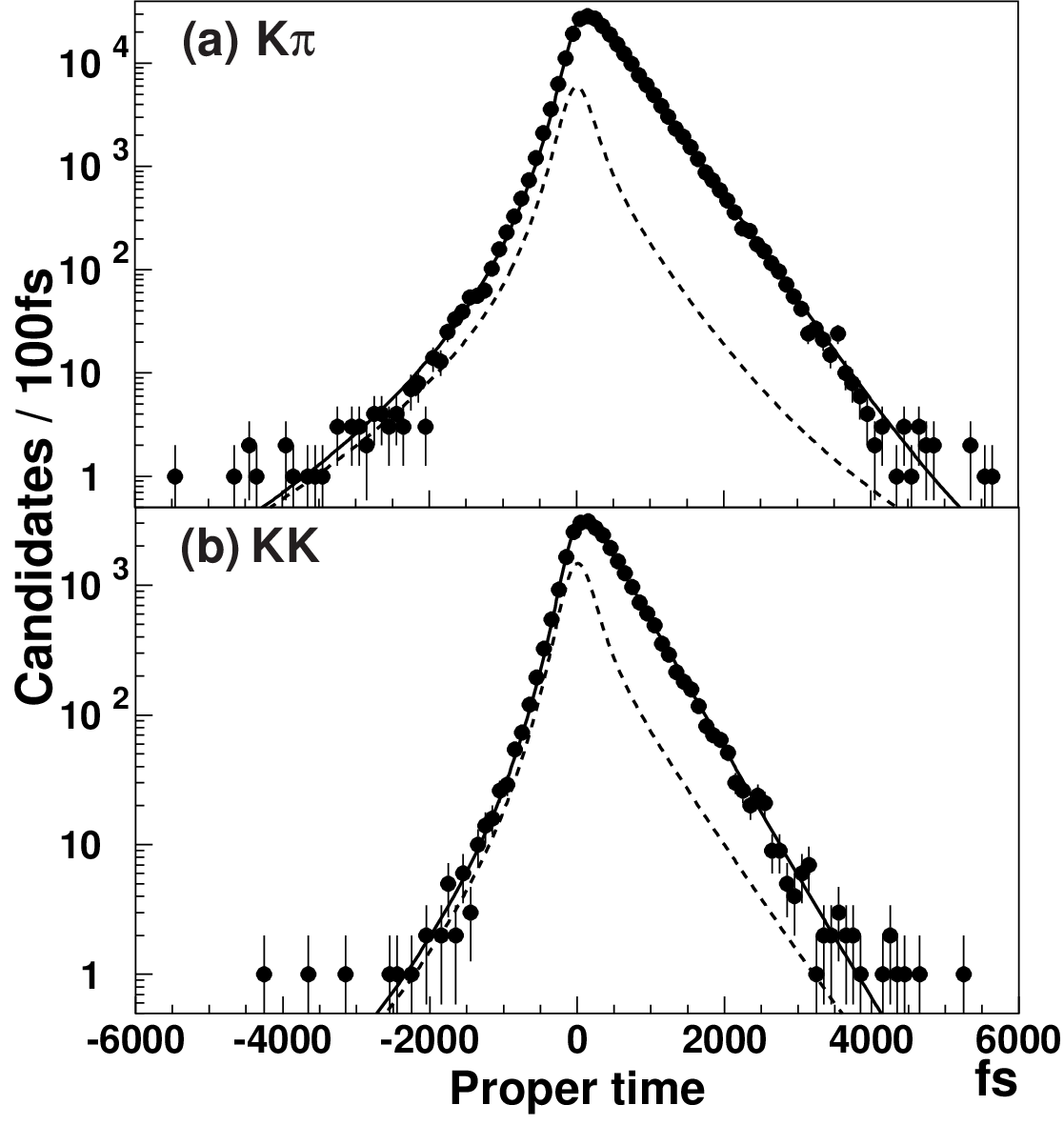}}
\end{center}
\begin{center}
\caption{Proper-time distributions and fit results for the decay modes (a) $\dzkpi$
and (b) $\dzkk$ in the \Dz\ mass signal region.
The solid line is the result of the fit.
The dotted line indicates the background contribution.}
\label{fig:proper-time-sig}
\end{center}
\end{figure}
\begin{figure}
\begin{center}
\resizebox{0.40\textwidth}{!}{\includegraphics{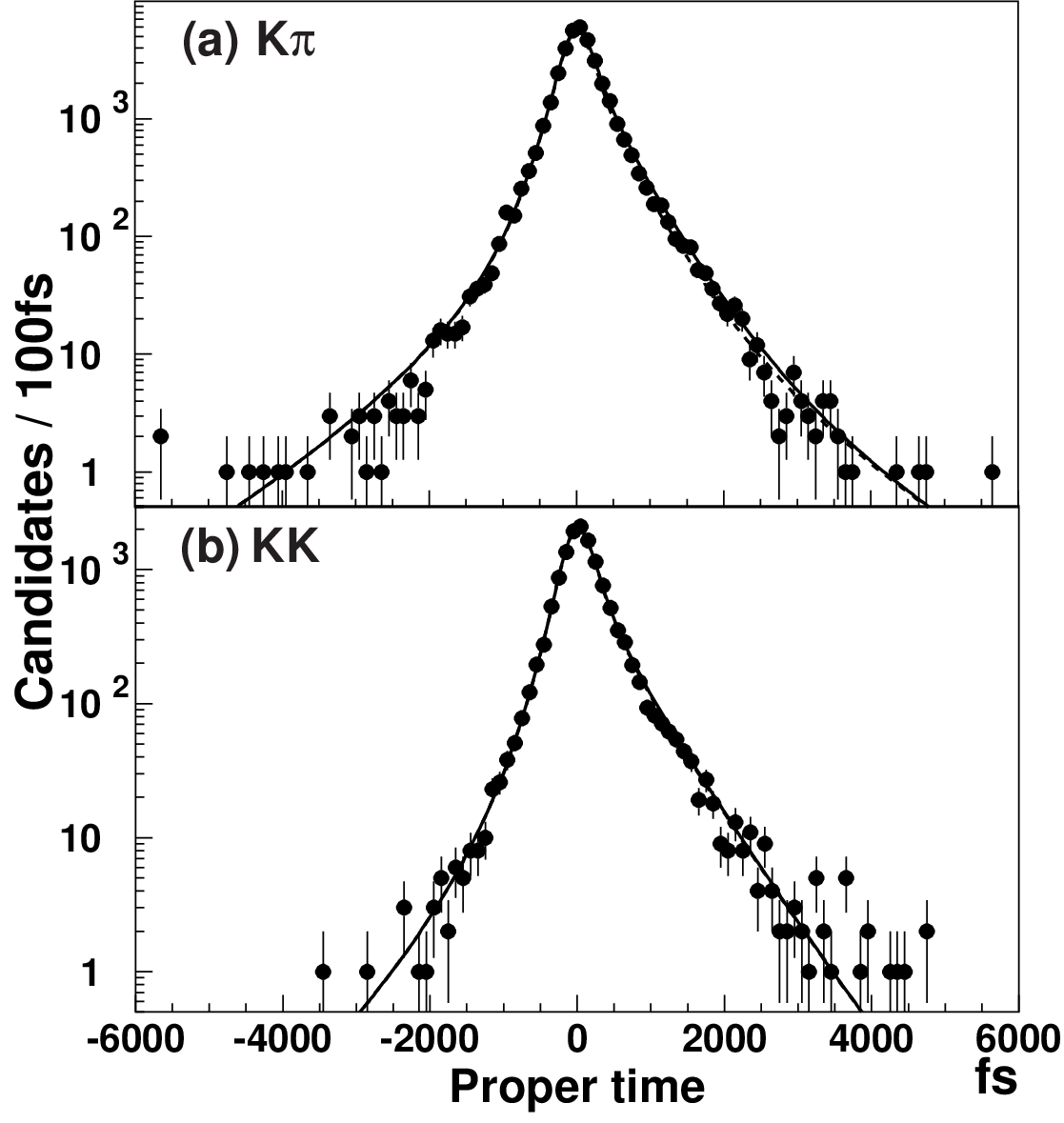}}
\end{center}
\begin{center}
\caption{Proper-time distributions and fit results for the decay modes (a) $\dzkpi$
and (b) $\dzkk$ in the \Dz\ mass background-dominated region.}
\label{fig:proper-time-bg}
\end{center}
\end{figure}

A simultaneous fit is performed to all the $D^0$ candidates
contained in the mass region within $40~{\rm MeV}/c^2$ of the mean $D^0$ mass.
This wide range includes both the signal~($<\pm3\sigma$) and the
background-dominated~($>\pm3\sigma$) regions.
Figure~\ref{fig:proper-time-sig} shows the results of the fit in the signal region.
The solid lines show the fit and the dotted lines show
the background contribution in the fit.
Figure~\ref{fig:proper-time-bg} shows the results of the fit in the background-dominated
region and demonstrates that the background shape is well reproduced in the fit.
The fit yields $f_{\rm tail}\sim 0.17$, $S\sim 0.84$, and  $S_{\rm tail}\sim 1.75$ resulting in an average
proper-time resolution of 215~fs for signal, while $f_{\rm tail}^{\rm BG}\sim 0.06$, $S_{\rm BG}\sim 1.05$ and
$S_{\rm tail}^{\rm BG}\sim 4$ for background.
These values are found to be nearly identical for the $K^-\pi^+$ and $K^-K^+$ decays.

We obtain  $y_{CP}=(-0.2\pm 1.0)\times10^{-2}$ and $\tau(K^- \pi^+)=416.2\pm 1.1$~fs from the fit.
The fit results are corrected for small biases found in a large sample of MC events, where
the reconstructed proper-time is smaller than the generated value. 
The difference diminishes when the effects of multiple
scattering, decay in flight, or hadronic interaction are turned off  in the MC simulation~\cite{Geant}.
The bias is found to be decay-mode dependent, 
$-1.5 \pm 0.6$~fs for $\dzkpi$ and
$-2.7 \pm 1.2$~fs for $\dzkk$, resulting in a correction for $\ycp$ of $(-0.3\pm 0.3)\times10^{-2}$.
Here errors are due to the MC
sample statistics and included as a systematic error in the final result. 
With this correction to the fit we obtain the result $y_{CP}=(-0.5\pm 1.0)\times10^{-2}$.

The systematic uncertainties for $y_{CP}$ are listed in Table~\ref{table:systematics}. 
Because $\ycp$ is measured as a ratio of two lifetimes,
many correlated systematic uncertainties in the reconstructed decay length cancel; as a
result, errors in decay vertex measurement, uncertainties in the
IP profile and the uncertainty of the reconstructed $D^0$ momentum
vector all make negligible contributions to the uncertainty of $\ycp$.
A MC simulation study shows that the fit yields a signal lifetime slightly longer than the generated one due to the presence of the background.
This bias is common to the $K^-\pi^+$ and $K^-K^+$ decays in sign and magnitude~($\sim$$+$3~fs) and, therefore,
has no effect on the value of $y_{CP}.$

A difference in the signal purity between the $K^-\pi^+$ and $K^-K^+$ decays may result in a bias in $y_{CP}$.
We estimate this uncertainty by repeating the analysis for $D^0$ samples 
obtained by varying requirements for particle identification and the fit quality of the $D^0$ decay vertex,
both of which are very effective for suppressing background.
The systematic uncertainty due to the background proper-time distribution is estimated by
varying the $D^0$ mass range used in the fit from the nominal 
$\pm 40$~\Mevcsq\ to $\pm 35$~\Mevcsq\ and to $\pm 45$~\Mevcsq,
and by comparing the results with different parametrizations.

The systematic error originating from a correlation between 
the  proper-time measurement and the measured $D^0$ mass
is estimated by a MC simulation study. 
We estimate the systematic uncertainty due to $D^0$ candidates with large proper times~($t>10\tau_{D^0}$)
by varying the $t$ range for the fit and taking the maximum excursion to be the systematic error.
Also included are the effects of the systematic uncertainties due to the statistical uncertainty of the signal probability $f_{\rm SIG}$ 
and the error of the world average value of the $D^0$ mass~\cite{pdg}.
The total systematic error is calculated by taking a quadratic sum of all contributions and
is $^{+0.007}_{-0.008}$.

In summary, we have measured the \DzDzbar\ mixing parameter $\ycp$,
using \intlycp~\Fb\ of data collected
near the $\Upsilon$(4$S$) resonance, to be
$$y_{CP}=\yresult.$$ 
This corresponds to a 95\% confidence interval \Ylimit.
The result is consistent with zero and
the Standard Model expectation that $\ycp$ is small.

We wish to thank the KEKB accelerator group
for the excellent
operation of the KEKB accelerator.
We acknowledge support from the Ministry of Education, Culture, Sports, Science and
Technology of Japan and
the Japan Society for the Promotion of Science;
the Australian Research Council and the Australian Department of Industry,
Science and Resources;
the Department of Science and Technology of India;
the BK21 program of the Ministry of Education of Korea and
the SRC program of the Korea Science and Engineering Foundation;
the Polish State Committee for Scientific Research
under contract No.2P03B 17017;
the Ministry of Science and Technology of Russian Federation;
the National Science Council and the Ministry of Education of Taiwan;
and the U.S. Department of Energy.

\newpage

\begin{table}
\begin{center}
\caption{Summary of systematic errors on the $\ycp$ measurement.}
\label{table:systematics}
\end{center}
\begin{center}
\begin{tabular}{lr}
Source                               & Systematic error~($10^{-2}$) \\ \hline \\
Reconstruction bias correction       & $\pm0.3$ \vspace{0.8mm} \\
Decay vertex error                   & negligible \vspace{0.8mm} \\
IP profile                           & negligible \vspace{0.8mm} \\
$D^0$ momentum error                 & negligible \vspace{0.8mm} \\
Particle identification              & $\pm0.5$ \vspace{0.8mm} \\
Decay vertex quality                 & $^{+0.1}_{-0.4}$ \vspace{0.8mm} \\
Background $t$ distribution          & $^{+0.2}_{-0.1}$ \vspace{0.8mm} \\
$D^0$ mass --  $t$ correlation       & $\pm0.3$ \vspace{0.8mm} \\
Large proper times                   & $\pm0.2$ \vspace{0.8mm} \\
Signal probability $f^i_{\rm SIG}$   & $\pm0.1$ \vspace{0.8mm} \\
World average $D^0$ mass             & $\pm0.1$ \vspace{0.8mm} \\ \hline \\
Total                                & $^{+0.7}_{-0.8}$ \\
\end{tabular}
\end{center}
\end{table}



\begin{references}
\bibitem{bias-exp}{
These distributions are not single exponentials in general.
Assuming a mixing parameter value $y <10^{-2}$, the resulting bias is 
less than a few percent of $y$ itself, and therefore negligible.
}
\bibitem{CC}{
	Charge-conjugate modes are implied throughout this Letter.
}
\bibitem{IP}{
	When a fill contains several runs, the beam position is recomputed
	for each run. In the case of very long runs, the beam position
	is also recalculated every $\sim$60,000 events.
}
\bibitem{sign}{
	Negative values of the decay length $\ell$ can occur due to finite vertex resolutions,
	as well as poorly-reconstructed vertices.
}
\end{references}
\end{document}